# THRESHOLD EXTINCTION IN FOOD WEBS

Bellingeri Michele

Department of Environmental Sciences, University of Parma, Viale Usberti, 33/A 43100 Parma, Italy

## ABSTRACT

Understanding how an extinction event affects ecosystem is fundamental to biodiversity conservation. For this reason, food web response to species loss has been investigated in several ways in the last years. Several studies focused on secondary extinction due to biodiversity loss in a bottom-up perspective using in-silico extinction experiments in which a single species is removed at each step and the number of secondary extinctions is recorded. In these binary simulations a species goes secondarily extinct if it loses all its resource species, that is, when the energy intake is zero. This pure topological statement represents the best case scenario. In fact a consumer species could go extinct losing a certain fraction of the energy intake and the response of quantitative food webs to node loss could be very different with respect to simple binary predictions. The goal of this paper is to analyze how patterns of secondary extinctions change when higher species sensitivity are included in the analyses. In particular, we explored how food web secondary extinction, triggered by the removal of most connected nodes, varies as a function of the energy intake threshold assumed as the minimum needed for species persistence. As we will show, a very low increase of energy intake threshold stimulates a disproportionate growth of secondary extinction.

**Key words:** secondary extinction; food web robustness; in silico experiment; threshold extinction; species sensitivity.

## INTRODUCTION

Food webs describe "who eats whom" in ecosystems and they are from long time a central topic of ecology research (MacArthur 1955; May 1972; Pimm 1980; Dunne 2006; May 2006; Jordan et al. 2002; Montoya et al. 2006). Understanding how an extinction event affects ecological systems is highly relevant to biological conservation and, for this reason, food web response to species loss has been investigated in several ways (Dunne et al. 2002, 2004; Dunne and Williams 2009; Sole and Montoya 2001; Allesina and Bodini 2004; Allesina et al. 2009; Estrada 2007; Allesina and Pascual 2009).

The *rationale* indicates food webs as error resistant, i.e. food webs show low sensitivity to random removal of nodes; on the other hand, these systems are attack prone, that is, a considerable damage may be caused when removing the key species.

There are several types of removal criteria: we may remove the most connected nodes considering their connectance (Dunne et al. 2002, 2004; Sole and Montoya 2001; Allesina et al. 2009), or we may disconnect the bottlenecks if we are looking for their dominance (Allesina and Bodini 2004; Allesina et al. 2006) or expansibility (Estrada 2007). These processes test food webs stability by a common approach: they consider how the loss of species cascades into the further loss of biodiversity.

Moreover, almost all studies quoted above analysed qualitative-binary food webs that consider alimentary interactions as presence-absence. In these binary webs, all edges have the same magnitude, that is the amount of energy passing from one species to another is not specified. In reality, link strength can differ hugely (Banašek-Richter et al. 2009) and do not consider the energy extent allocated among species could hide the real food web response to species loss.

Only a recent analysis has investigated food web response to species loss accounting for link strength (Allesina et al. 2006). This work shows how secondary extinction grows faster in food web dominance context when weaker links are removed from the networks.

In a pure topological approach, the criterion for secondary extinction is simple: a species goes secondarily extinct when it has lost all its prey. This also means that a species goes extinct when no energy enters into it. This is the underlying statement supporting all binary models to forecast secondary extinction in food webs.

The claim that a taxon goes extinct when it loses 100% of energy intake is the best case scenario. In fact a consumer could be damaged enough to be pushed to extinction before the total energy intake has vanished (Bodini et al. 2009). A more sensible species could go extinct when it loses a certain fraction of energy intake, e.g. 70%. For this reason, the species response to energy intake reduction appears as a fundamental pattern to understand food web reaction to node loss. Nonetheless, such question is still largely unexplored.

The main objective of this paper is to investigate how patterns of secondary extinctions change when species sensitivity and links magnitude are included in the analyses. In this study we analysed a set of 18 weighted food webs throughout a new extinction scenario that considers species sensitivity to energy intake decrease. In particular, we explored how food web secondary extinction to removal of most connected nodes (Dunne et al. 2002, 2004; Sole and Montoya 2001) vary as a function of the energy intake threshold assumed as the minimum needed for species persistence. As we will show, in almost all extinction scenarios, a very low increase of energy intake threshold induces a disproportionate growth of system fragility.

This discovery unveils a possible underestimation of secondary extinction in biological community made in qualitative predictions. At the same time, we established a bridge between the classical binary food web analyses and new quantitative methods able to improve ecological networks study.

# METHODS

In this work we approach the secondary extinction in ecosystems using quantitative-weighted food webs instead of the typical empirical binary networks used in other studies (Sole and Montoya 2001; Dunne et al., 2002, 2004; Allesina and Bodini 2004; Allesina et al. 2009). We analysed the food webs of 18 ecosystems of various size (min size S=23, max size S= 248). We chose only webs with species richness S > 14 to avoid bias due to small web size (Bersier and Sugihara 1997). They were previously investigated as ecological flow networks, that are graphs of ecosystems in which the magnitude of trophic transfers from prey to predators is known (Ulanowicz 1986).

Two web sites provided the data about these ecosystems. One set (8 graphs) is made available in the project site ATLSS (Across Trophic Level System Simulation, http://www.cbl.umces.edu/~atlss/). Others 8 models are in the Prof. Ulanowicz's web page (http://www.cbl.umces.edu/~ulan/ntwk/network.html).

The former database includes high quality data corresponding to 4 ecosystems over two seasonal steps (wet and dry season), while the latter was a more heterogeneous database (e.g., the 4 models of the Chesapeake Bay describe different geographical areas in summer season – Upper, Middle and Lower – and levels of compartment aggregation – Chesapeake Mesohaline Ecosystem). At the same web page are available others four networks: Mondego Estuary, St. Marks river, Lake Michigan, Final Narraganset bay. Finally the Caribbean reef food web was analyzed in Bascompte et al. (2005) and the network of the small mountain Lake Santo is described in Bondavalli et al. (2006). According to food web theory, we have excluded non-living nodes from the networks. The complete list with a coarse description of the 18 food webs is in Table 1.

In a classic extinction scenario, a species is considered extinct when it loses all its resources. From a topological point of view, this means that a node becomes extinct if it does not have qualitative incoming links. In other words, a species goes extinct after the entire energy intake has been lost.

If we assign a threshold value of consumer energy intake below which a species is damaged enough to go extinct, and we indicate this threshold by $v$, in classic extinction scenario $v$ is implicitly assumed equal to 0. In other words, a species goes extinct when the inflow energy is null (Dunne et al. 2002, 2004; Dunne and Williams 2009; Sole and Montoya 2001; Allesina and Bodini 2004; Allesina et al. 2009).

In a more general way, we indicate with $e(i)$ the current inflow energy to species $i$ after nodes removal and in our model a species $i$ goes extinct if:

$$e(i) \leq v \qquad (1)$$

that is when the inflow energy into the node $i$ is less or equal to the minimum necessary to species survival. From now on we name $v$ Threshold Extinction. Current inflow energy $e(i)$ is normalized by the starting inflow energy (i.e. inflow energy to species $i$ before any removal), for this $e(i)$ is a value within the interval (0,1). Clearly, the higher the Threshold Extinction is, i.e. the more energy is necessary to species survival, the more sensitive the species will be to energy intake decrease. For the same reason, the higher the Threshold Extinction is, the more food webs should be sensitive to node loss.

In binary extinction scenario, the nodes were removed at each step, and the amount of species with no resources was assessed (Dunne et. al 2002, 2004; Sole and Montoya 2001; Allesina et al. 2009; Allesina and Pascual 2009). On the contrary, in our extinction scenario we removed nodes, measuring how many species at each step had current inflow energy lower or equal to Threshold Extinction. According to the condition (1), these taxa will go extinct. We produced a set of extinction scenario in ascending order of Threshold

Extinction, i.e. from 0 to 1 increasing $v$ of 10% at every step. For $v$=0, we obtained a binary extinction scenario (Dunne et al. 2002, 2004; Sole and Montoya 2001) in which species go extinct when the energy intake is zero, i.e. all resources have been lost.

When the setting $v$=0.1, we considered a species extinct if the energy intake is equal or lower than 10% of the original diet, that is when the 90% of inflow is lost. When setting $v$=0.2, we considered a species extinct when energy intake becomes equal or lower than 20%, that is when the 80% of the starting inflow is lost. We repeated this procedure by increasing species sensitivity up to $v$=1.

We performed extinction scenario by two well known removal criteria (Dunne et al. 2002, 2004): 1) removing the most connected species at each step (i.e. species with majour number of connections); 2) removing the most outgoing connected species at each step (i.e. species with largest amount of connections coming out). For each food web we computed the difference between binary secondary extinction magnitude (i.e. extinction scenario by $v$=0) and secondary extinction made by following quantitative Threshold Extinction. We name this distance "extinction gap" and it represent the growth of food web sensitivity as a function of Threshold Extinction. Finally, we used linear regressions to examine the relationship between the increase in secondary extinction and two measures of food web complexity: species richness (S ) and connectance ($C= L/S^2$).

**Table 1**. Ecosystems and their food web statistics. S= number of species; L = number of links, L/S = Linkage density; C = Connectance; Keys indicates the food web label in the following figures. The keys DRY and WET identify food webs of the same ecosystem referring to dry and wet season respectively.

| FOOD WEBS | S | L | L/S | C | Keys |
|---|---|---|---|---|---|
| LAKE SANTO | 23 | 140 | 6.08 | 0.26 | a |
| FINAL NARRAGANSETT BAY | 31 | 113 | 3.65 | 0.12 | b |
| CHEASEPEAKE LOWER | 31 | 57 | 1.84 | 0.06 | c |
| CHEASEPEAKE MIDDLE | 31 | 77 | 2.48 | 0.08 | d |
| CHEASEPEAKE UPPER | 31 | 83 | 2.68 | 0.09 | e |
| CHEASEPEAKE MESOHALINE | 33 | 121 | 3.66 | 0.11 | f |
| LAKE MICHIGAN | 35 | 130 | 3.71 | 0.11 | g |
| MONDEGO | 42 | 279 | 6.64 | 0.16 | h |
| ST. MARK RIVER | 48 | 219 | 4.56 | 0.1 | i |
| EVERGLADES GRAMINOIDS DRY | 63 | 617 | 9.79 | 0.16 | l |
| EVERGLADES GRAMINOIDS WET | 63 | 576 | 9.14 | 0.15 | m |
| CYPRESS WETLAND DRY | 65 | 448 | 6.89 | 0.11 | n |
| CYPRESS WETLAND WET | 65 | 439 | 6.75 | 0.10 | o |
| MANGROVE DRY | 91 | 1149 | 12.62 | 0.14 | p |
| MANGROVE WET | 91 | 1151 | 12.65 | 0.14 | q |
| FLORIDA BAY DRY | 123 | 1799 | 14.76 | 0,13 | r |
| FLORIDA BAY WET | 123 | 1767 | 14.5 | 0.12 | s |
| CARIBBEAN | 248 | 3264 | 13.16 | 0.05 | t |

# RESULTS

As we expected, secondary extinction increases as a function of Threshold Extinction  in all our simulations produced. The higher the minimum energy intake to species survival is, the more likely secondary extinction by the same removal criteria is. Secondary extinction outcome are represented as in Allesina et al. (2009). See figure 2 and 3 for a representation of the secondary extinction outputs.

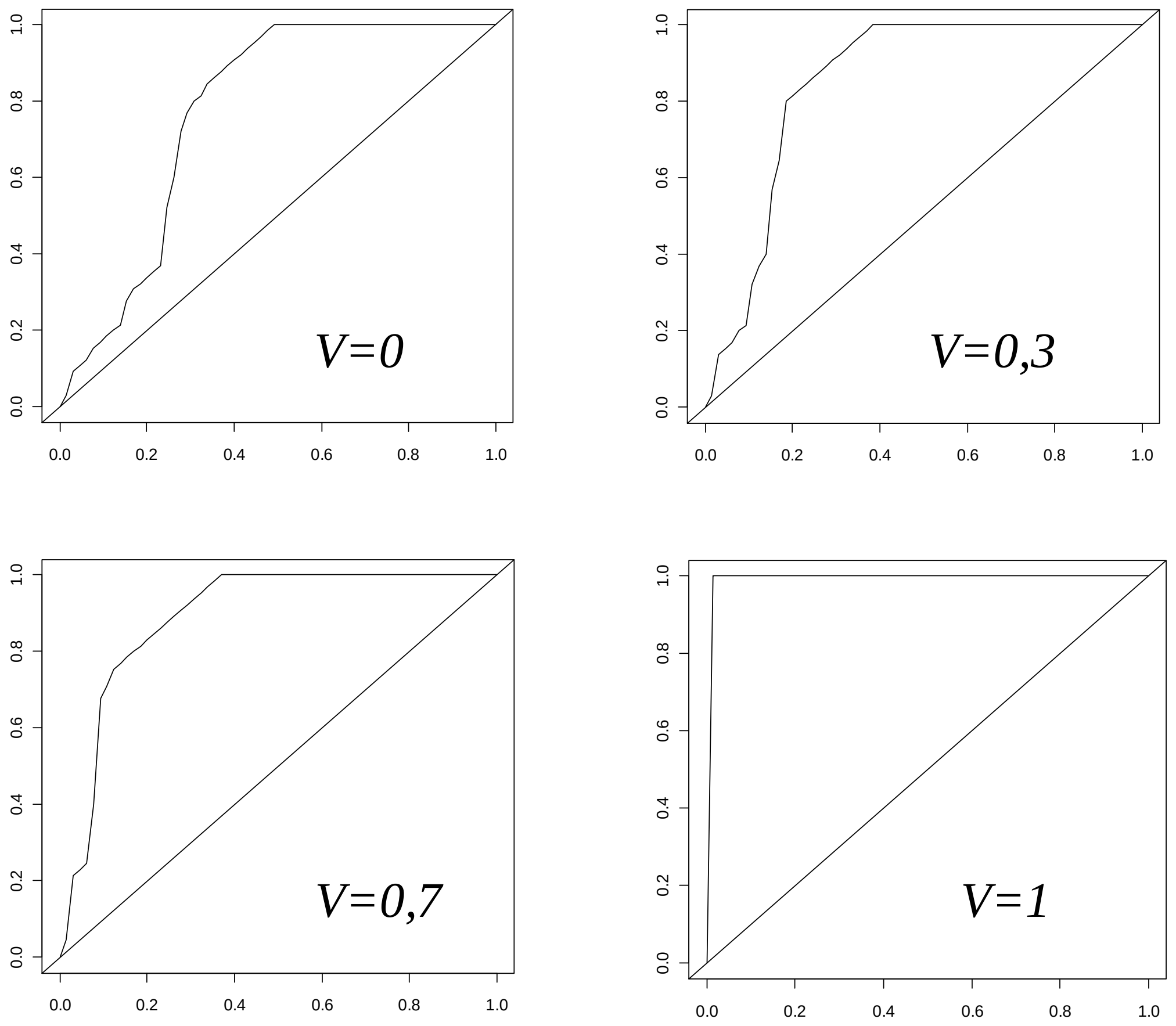

**Figure 2**: most connected removal extinction scenario as a function of $v$ in Cypress wet food web S=65. Percent of nodes removed (x axis) and percent of nodes extinct, i.e. nodes removed plus secondary extinction (y axis). The bisector line means no secondary extinction scenario. In the top left corner of the figure, $v$=0 is a classic qualitative extinction scenario in which species go extinct losing all resources, as in Dunne et al. (2002). When increasing $v$=0.3, we can see an higher shape of extinction curve indicating larger secondary extinction. For $v$=0.7 we see a further secondary extinction, since we reach the limiting and trivial extinction scenario in which $v$=1. In this simulation all species go extinct after first removal.

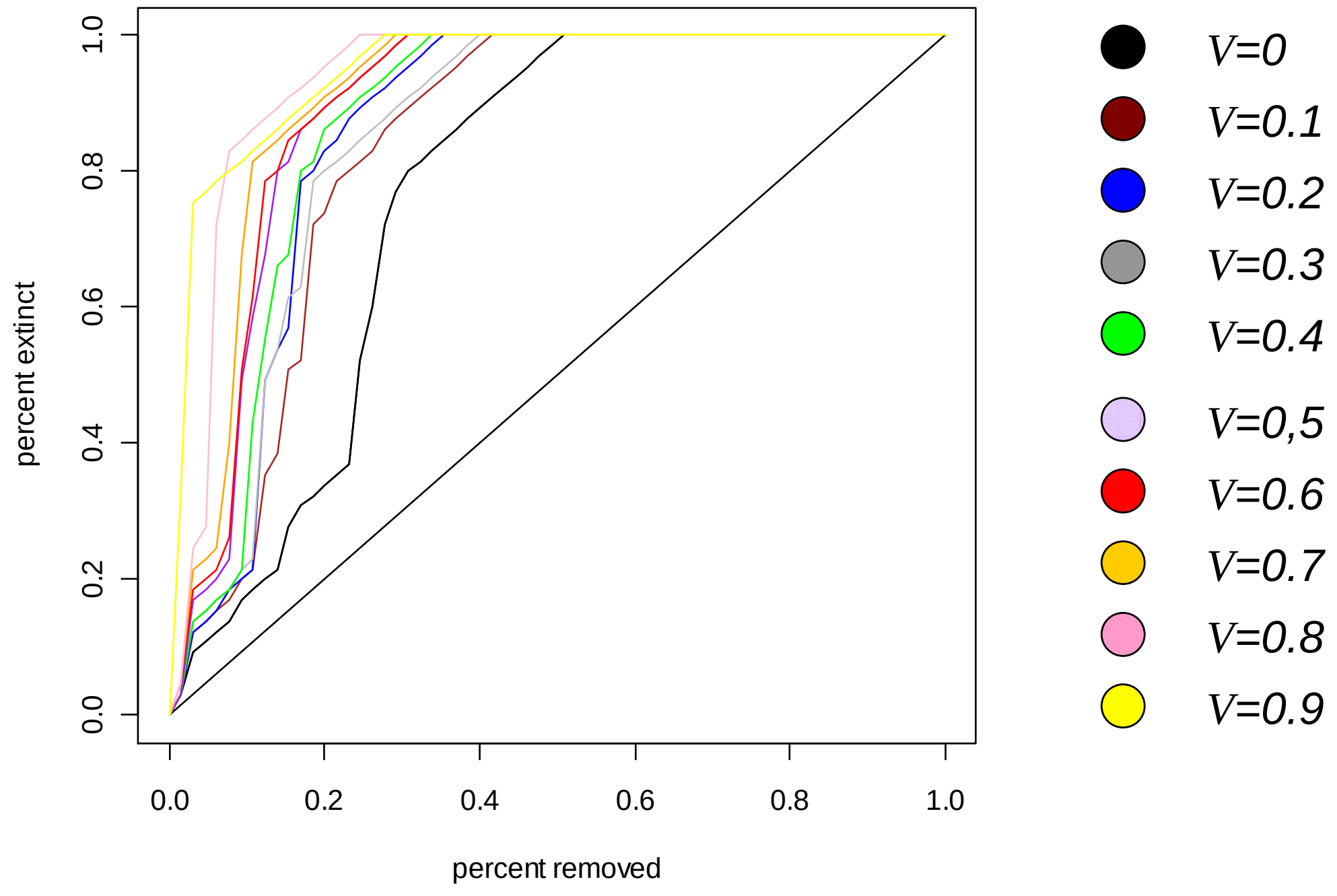

**Figure 3**: most connected removal extinction scenario for the entire increasing set of Threshold Extinction. On the right side of the figure we can see colours corresponding to a $v$ value of the curve. The black curve represents the $v$=0 extinction scenario, the brown curve represents extinction scenario for $v$=0.1, and so forth up to $v$=0.9. The $v$=1 extinction scenario has been omitted. The positive relation among Threshold Extinction and secondary extinction magnitude is displayed by the growing derivative of the curves. This means that the system robustness decreases when we take into account an higher species sensitivity.

If secondary extinction is expected to rise due to an increase in Threshold Extinction, the growth rate is surprisingly high. Extinction scenario outputs show a system response that changes abruptly by a minimum increase of $v$. In almost all extinction scenarios, we notice a sharply increase in secondary extinction when we move from $v$=0 to $v$=0.1 or $v$=0.2. In other words, system fragility rises very fast when applying a slightly higher species sensitivity. Figure 4 shows this transition of food web response as a function of Threshold Extinction in five networks for both removal criteria. In this figure we can see that the main mismatch from curves occurs at the beginning of the assumed Threshold Extinction set (i.e. from $v$=0 to $v$=0.2). Since a minimum Threshold Extinction increase produces a sizeable rise of secondary extinction we concentrate the following analysis on first steps of $v$, that is $v$=0.1 and $v$=0.2.

In order to quantify the extinction gap we have calculated the difference between $v$=0 and $v$=0.1 extinction curves, then between $v$=0 and $v$=0.2.

In doing this, we discovered another weighty scenario: a maximum extinction gap emerges at the first steps of the removal sequence. That is, the largest underestimate in terms of secondary extinction occurs after few species removals. For example, in Cypress wet food web an extinction gap equal to 45% emerges when removing less than 10% of species during the transition from $\nu$=0 to $\nu$=0.1. In Caribbean food web outputs, as we can see in figure 4e and 4l, it is enough to remove 10% of species to assist to 70% extinction underestimate moving from $\nu$=0 to $\nu$=0.2 for both removal criteria. Since a large Caribbean food web has 248 species, about 170 species become extinct, in case we assume species sensitivity a minimum higher than the binary one. The entire set of outputs is in Appendix.

Eventually, secondary extinction gap increases significantly with increasing species richness of the food web. On the contrary, we found the secondary extinction increase did not vary significantly with connectance. Linear regressions are in Table 2.

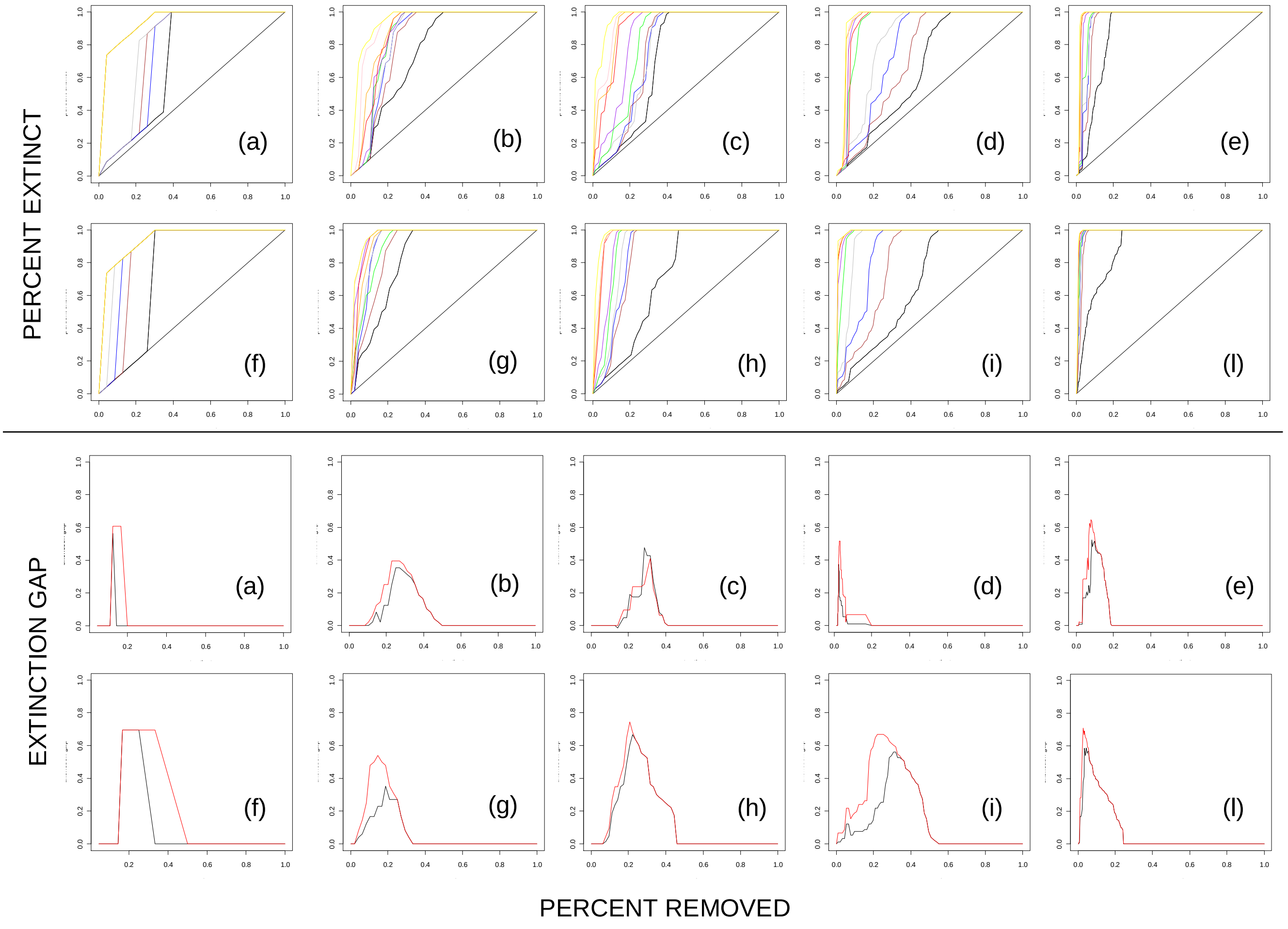

**Figure 4**: five extinction scenario outputs from eighteen food webs analysed. The figures in the first row (a-e) show most connected removal extinction scenarios. The figures in the second row (f-l) display the extinction scenario for most outgoing connected removal criteria. Each figure in the third and fourth row indicates the extinction percentage gap of the corresponding extinction scenario shown in the first two rows. The black curve represents the difference in terms of secondary extinction moving from $\nu$=0 to $\nu$=0.1; the red curve represents the difference in terms of secondary extinction moving from $\nu$=0 to $\nu$=0.2. Food web keys are in table 1.

**Table 2**: Simple linear regressions of the secondary extinction underestimate as a function of S and C. Significant regressions are shown in bold.

| | Species Richness (*S*) | | | Connectance (*C*) | | |
|---|---|---|---|---|---|---|
| Deletion sequence | Slope | *P* | $r^2$ | Slope | *P* | $r^2$ |
| Most connected $\nu$=0.1 | **0.002** | **< 0.001** | **0.54** | -0.06068 | 0.93 | 0.0004 |
| Most connected $\nu$=0.2 | **0.0022** | **< 0.001** | **0.5** | -0.5138 | 0.56 | 0.0022 |
| Most outgoing connected $\nu$=0.1 | **0.002** | **0.001** | **0.35** | 0.06634 | 0.94 | 0.0004 |
| Most outgoing connected $\nu$=0.2 | **0.0017** | **0.003** | **0.43** | -0.01052 | 0.99 | 0.0001 |

# DISCUSSION

The accelerating and worldwide loss of biodiversity is a matter of growing concern. Owing to the interdependences among species in ecosystems, such losses of species might in turn unleash cascades of secondary extinctions (Pimm 1980; Dunne et al. 2002; Montoya et al. 2006; Dunne et al. 2009; Ebenman 2011, Allesina and Bodini 2004). For this reason, in the last years several studies have analyzed the response of binary food webs to species loss in terms of secondary extinction (Dunne et al. 2002, 2004; Dunne and Williams 2009; Sole and Montoya 2001; Allesina and Bodini 2004; Allesina and Pascual 2009). Our findings show how secondary extinction in food webs grows tuning an increasing set of Threshold Extinction. Just assuming a bit higher threshold we found an outstanding increase of secondary extinction magnitude after few species removals. It is noteworthy to focus on the $\nu$=0.1 extinction scenario. In many food webs analyzed tuning $\nu$=0.1 is enough to produce a secondary extinction increase equal to 40% or more. In other words, when we assume that a consumer goes extinct following up the loss of 90% of energy intake rather than 100%, as in binary models, food webs sensitivity to targeted species loss quickly rises. Such outcomes indicate how binary food webs forecast could underestimate secondary extinction magnitude in the case real consumer sensitivity is higher than the simple qualitative statement.

Allesina et al. (2006) is the first quantitative investigation in which they applied the dominator tree model in eight weighted food webs. Firstly, the analysis consisted in removing links below an increasing threshold of magnitude, and secondly in building the dominator tree associated to the remaining structure. A sharply increase of secondary extinction was discovered just imposing links threshold equal to 15%. In other words, a great amount of species emerged as dominated by other nodes after the weaker links had been removed. Although performed by different methods, ours analyses confirm Allesina et al. (2006) results.

Regarding the diversity-stability relationship (Mc Cann 2000), we discovered a positive linear fit between food web size and maximum extinction gap passing from $\nu$=0 and $\nu$=0.1-0.2 outcomes. That is, larger systems display higher secondary extinction gap and food webs sensitivity as a function of Threshold Extinction follows species richness. In addiction, the rise of secondary extinction with Threshold Extinction is not related with connectance. Thus, neither connectance or size seems to be able to mitigate the increase in secondary extinction when we assume higher Threshold Extinction for species. Ours findings differs from binary predictions in which food webs display increasing robustness to loss of highly connected species with connectance (Dunne et al. 2002, 2004). The structurally stabilizing role of increased connectance found in binary models seems not to be preserved in the case we analyze food webs resistance from a quantitative point of view.

# CONCLUSION

In this papers we has shed light on the possible underestimation of the secondary extinction carried out in binary-qualitative analyses. In addiction to that, we demonstrated how species sensitivity pattern is central topic to understand food webs response under perturbation. Thus system stability is not only affected by the topological structure, but it is strictly related to the robustness of its components (species, taxa, comparts). It is necessary to remark that the basic assumption of our scenario postulates all species as equal robustness (i.e. all species have the same Threshold Extinction), but within a biological community, species could have several sensitivity (Ebenman 2011, Bodini et al. 2009). For example autotrophs could be more resistant to energy intake decrease than apical predator. In conclusion, the methods presented in this work could be further applied to investigate how Threshold Extinction distribution affects food web stability.